\documentclass[twocolumn,showpacs,preprintnumbers,amsmath,amssymb,superscriptaddress]{revtex4}

\usepackage{graphicx}
\usepackage{dcolumn}
\usepackage{bm}

\begin{document}


\title{Chiral Magnetic Effect in a Lattice Model}

\author{Bo  Feng}
\affiliation{
School of Physics, Huazhong University of Science and Technology, Wuhan 430074, China
}

\author{De-fu Hou}
\affiliation{
Institute of Particle Physics, Huazhong Normal University, Wuhan 430079, China}

\author{Hui Liu}
\affiliation{
Physics Department, Jinan University, Guangzhou, China}
\author{Hai-cang Ren}
\affiliation{
Physics Department, The Rockefeller University, 1230 York Avenue, New York, New York 10021-6399, USA}
\affiliation{
Institute of Particle Physics, Huazhong Normal University, Wuhan 430079, China}
\author{Ping-ping Wu}
\affiliation{
School of Physics and Electronic Information Engineering, Henan Polytechnic University, Jiaozuo 454000, China}
\author{Yan Wu}
\affiliation{
School of Mathematice and Physics, China Geoscience University(Wuhan), Wuhan 430074, China }

\date{\today}

\begin{abstract}

We study analytically the one-loop contribution to the Chiral Magnetic Effect(CME) using lattice regularization with a Wilson fermion field. In the continuum limit, we find that the chiral magnetic 
current vanishes at nonzero temperature but emerges at zero temperature consistent with that found by Pauli-Villas regularization. For finite lattice size, however, the chiral magnetic current is 
nonvanishing at nonzero temperature. But the numerical vaule of the coefficient of CME current is very small compared with that extracted from the full QCD simulation for 
the same lattice parameters. The possibility of higher order corrections from QCD dynamics is also assessed.

\end{abstract}

\maketitle

\section{Introduction}

While the global $\cal P$ and $\cal CP$ violation is absent in Quantum Chromodynamics(QCD), which has been restricted to a stringent limit via the measurement of the electic 
dipole moment of the neutron, the possibility of their local violation in hot QCD matter has been proposed to be observable  in noncentral heavy ion experiment through the 
Chiral Magnetic Effect(CME)\cite{CME1,CME2,CME3}.  Such a violation also serves an important signal of the formation of quark-gluon plasma during the collision. Moreover, it has been realized 
recently that CME can also be implmented in condensed matter system-a (3+1)-dimensional Weyl semimetal\cite{Weylsemi1,Weylsemi2,Weylsemi3,Weylsemi4}, in which 
the quasiparticles in the vicinity of the band touching point are approximately massless chiral fermions with linear relation between their energies and momenta, 
and thus chiral anomaly that drives CME may show up\cite{NN1}.
   
One remarkable feature of non-Abelian gauge theories such as QCD is the existence of topologically inequivalent configurations of gauge fields characterized by different 
winding numbers. The transitions between these different sectors via instantons at zero temperature or via sphalerons at  finite temperature would induce a  
chirality imbalance via chiral anomaly\cite{tHooft1,tHooft2,JackiwRebbi,CallanGross}. While these topological charge fluctuations are not directly observable, it can induce observational signals in experiment in 
the presence of a strong magnetic field produced in noncentral heavy ion collisions. This magnetic field aligns the spin and momentum directions of the massless quarks 
parallel/antiparallel to it according to their chiralities.  Consequently, a seperation of electric charge and thus a vector current(CME current) is generated along the 
direction of magnetic field if there is imbalance between the left- and right-handed quarks(A new mechanism has found recently, in which the CME current can be generated 
without initial chirality imbalance by reconnections of magnetic flux\cite{Kharzeev}). Therefore, CME provides a very promising opportunity to observe the local $\cal P$ and $\cal CP$-odd 
effects and to probe the nontrivial topologies in hot QCD matter. There are already analyses on experimental data showing the signals expected from CME\cite{CMEinRHIC1,CMEinRHIC2,CMEinAlice} and further 
investigations to scrutinize the backgrounds and/or to rule out other explanations are in progress\cite{CMEtaskforce}. 
 
There have been a large body of literatures on the theoretical aspect of CME. While a systematic calculation integrating the creation process of the chirality imbalance and the 
subsequent electric current generated in response to a transient strong magnetic field is analytically difficult, a common approach is to model a spot of chirality imbalance 
with an equilibrium ensemble at a nonzero chemical potential $\mu_5$. A classical form of CME current(for one color and one flavor with a unit electric charge $e$) in a constant 
magnetic field $\bf B$
\begin{equation}
{\bf J}=C{e^2}\mu_5 {\bf B}
\label{classical}
\end{equation}
with $C=1/(2\pi^2)\simeq 0.05$ is then obtained throughout the literatures without a UV regularization. Later works, however, revealed a number of subtleties behind this approach. 
It was argued 
in \cite{rubakov} and was later confirmed with a one-loop calculation by Pauli-Villars (PV) regularization \cite{HouRen} that there is no chiral magnetic current at  nonzero 
temperature unless a magnetic helicity term is added to the axial charge. Relaxing the constancy condition of $\mu_5$, 
the static and homogeneous limit of $\mu_5$ turns out to be order dependent\cite{HouRen}. If the time dependence of a spatially homogeneous $\mu_5$ is 
switched off, the classical CME shows up without adding the magnetic helicity. If the homogeneity limit is taken for a static axial chamical potential the effect 
disappears, just like the PV-regularized one-loop result. The consequences of both orders of limit appears robust to all orders of perturbation if the infrared divergence 
of QCD can be eliminated.  
    
Lattice simulation is a powerful tool to explore nonperturbative aspects of non-Abelian gauge theories and has been employed to investigate the interplay between the 
topological transitions and the subsequent local $\cal P$ and $\cal CP$ violations\cite{Buividovich1,Abramczyk,Yamamoto,Yamamoto1,Buividovich2,Schafer}.  
The axial chemical potential $\mu_5$, unlike the baryon one, does not give rise to sign problem since it always makes the fermion determinant positive definite, making the 
lattice simulations of a thermal ensemble with an nonzero $\mu_5$ feasible. Several simulations along the line have been reported in the literature. In particular, the 
simulation of QCD with two flavors on a lattice of size $12^3\times 4$ with Wilson fermions revealed a chiral magnetic current which  
appears linear in both $\mu_5$ and $\bf B$, consistent qualitatively with the classical results \cite{Yamamoto} but with much smaller magnitude of $C(\simeq 0.013)$ in eq.(\ref{classical}). 
The nonzero chiral magnetic current seems in contradiction to null result obtained from the one-loop approximation under PV regularization and the present work is
addressing this difference.  



In this work, we calculated analytically the one-loop contribution on a lattice with a Wilson fermion field using the perturbative formulation developed in 
\cite{KarstenSmit}. In the continuum limit, we found that the chiral magnetic current
vanishes at  nonzero temperature but emerges at zero temperature, the same as the PV-regularized chiral magnetic current calculated in \cite{HouRen}. In another word, the 
lattice reglarization is completely equivalent to the PV regularization in this regard. Our result is consistent with the recent work in \cite{Buividovich2}, where the 
one-loop CME was studied numerically using the overlapping fermion formulation.

The rest of the paper is organized as follows: in Section II we present the fundamental formulas in lattice QCD and the one-loop calculation of CME current with a constant 
axial chemical potential at a nonzero temperature. In section III, the zero temperature chiral magnetic current is calculated and its relation to the chiral anomaly is 
discussed.  Section IV summarizes our work with a discussion on the possibility of the contribution from QCD dynamics to 
the nonzero chiral magnetic current extracted from the lattice simulation. 
Throughout the paper, we will work in Euclidean space with the four vector represented by $x^\mu=({\bf x},x_4), {q^\mu=({\bf q},i\omega_n})$ with $\omega_n$ the Matsubara 
frequency for bosons $\omega_n=2\pi nT$ and for fermions $\omega_n=(2n+1)\pi T$.

\section{General Formulae and One Loop Calculation}

We consider the following action with a single flavor of Wilson fermions in an external electromagnetic field on a hypercubic lattice
\begin{align}
\nonumber I=&-\sum_x\sum_{\mu}\frac{1}{2a}\left[{\bar{\psi}}(x)\left(\frac{1}{i}\gamma_\mu-r\right) U_\mu(x)\psi(x+a_\mu)\right.\\
\nonumber &-\left.{\bar{\psi}}(x+a_\mu)\left(\frac{1}{i}\gamma_\mu+r\right)  U_\mu^\dagger(x)\psi(x)
\right]\\
&-\sum_xM\bar{\psi}(x)\psi(x)
+\cdots \label{Wilsonaction}
\end{align}
with $a$ the lattice spacing and $M=m+4r/a$. The local gauge field is defined by the link variable
\begin{equation}
U_\mu(x)=\exp\left[iaev_\mu(x)\right]
\label{link}
\end{equation}
with $e>0$ the electromagntic coupling constant.  The momentum representation of the free fermion propagator  is
\begin{equation}
S(p)=a\left[\sum_{\mu}\gamma_\mu\sin ap_\mu +{\cal M}(ap) \right]^{-1}
\end{equation}
with the mass term
\begin{equation}
{\cal M}(ap)=aM-r\sum_{\mu}\cos ap_\mu
\end{equation}
and the momentum space is a Brillouin zone defined by $-\frac{\pi}{a}<p_\mu\le\frac{\pi}{a}$.
In chiral limit(m=0), the mass term becomes
 \begin{equation}
 {\cal M}_c(ap)=r\sum_{\mu}\left(1-\cos ap_\mu\right)
 \end{equation}

In the absence of the Wilson term, ${r}=0$, one finds the naive lattice fermion action. There are 16 fermion species of massless fermions in the continuum limit in accordance
with the Nielsen-Ninomiya theorem\cite{NN2} and the chiral symmetry holds exactly. For $r\neq 0$, the Wilson term lifts the masses of fifteen of them to $O\left(\frac{1}{a}\right)$
and thereby breaks the chiral symmetry explicitly. The Feynman rules for the perturbative expansion in $e$ can be found in\cite{KarstenSmit},
where the vertex functions are more complicated since there are vertices with more
than one gauge fields coming from the higher order terms of (\ref{link}). For this work, we need only the vertex function with one gauge field, which reads
\begin{equation}
V_\mu(p,q)=\gamma_\mu\cos \frac{1}{2}(ap_\mu+aq_\mu)+r\sin  \frac{1}{2}(ap_\mu+aq_\mu)\label{vertex}
\end{equation}

In the lattice model, the axial current can be constructed by introducing in the chiral invariant part of the action($m=r=0$) an external axial gauge field $a_\mu(x)$ as the author
of \cite{KarstenSmit} did, and the action becomes
\begin{widetext}
\begin{align}
\nonumber I=&\sum_x\left\{-\sum_{\mu}\frac{1}{2a}\left[{\bar{\psi}}(x)\frac{1}{i}\gamma_\mu \exp\{ia[ev_\mu(x)+\gamma_5 a_\mu(x)]\}\psi(x+a_\mu)-{\bar{\psi}}(x+a_\mu)\frac{1}{i}\gamma_\mu \exp\{-ia[ev_\mu(x)+\gamma_5 a_\mu(x)]\}\psi(x)
\right]\right.\\
&-\left.M\bar{\psi}(x)\psi(x)+\frac{r}{2a}\sum_\mu\left[{\bar{\psi}}(x) \exp[iaev_\mu(x)]\psi(x+a_\mu)+{\bar{\psi}}(x+a_\mu) \exp[-iaev_\mu(x)]\psi(x)\right]
\right\}+\cdots \label{yamamotoaction}
\end{align}
\end{widetext}
The axial chemical potential $\mu_5$ relevant for CME is then defined by replacing the axial gauge field with $a_\mu(x)\equiv\mu_5\delta_{\mu 4}$.
The fermion propagator becomes
\begin{equation}
S(p)=a\left[\sum_{\mu}\gamma_\mu\sin a\left(p_\mu+\gamma_5\mu_5\delta_{\mu 4} \right)+{\cal M}(ap) \right]^{-1}\label{axialpropagator}
\end{equation}
Upon a rescaling $\psi\rightarrow\psi/\sqrt{M}$ at $r=1$, the action (\ref{yamamotoaction}) is exactly the fermionic part of the action employed in\cite{Yamamoto} for the Monte Carlo
simulation, with $\kappa=1/(2Ma)$. The gluonic part, however, is irrelevent as far as the one-loop electric current is concerned.

The CME current is given as the linear response to the small external magnetic field\cite{HouRen,KharzeevandWarringa,Landsteiner}, i. e.
\begin{equation}
J_i(p)=-\Pi_{ij}(p)A_j(p)
\end{equation}
with $\Pi_{\mu\nu}(p)$ the photon polarization tensor and $A_\mu$ the external electromagnetic field. Here, the $\mu_5$ dependence of the photon polarization tensor
is implicit. The first term in the Taylor expansion of $\Pi_{\mu\nu}(p)$ in terms of $\mu_5$ corresponds to the usual triangle diagrams. The one-loop contribution
to the CME current had been studied extensively in the literature with continuum regulators. In this section, we shall calculate the one-loop
photon polarization(first order in $\mu_5$) using the lattice regularization, which is a natural and non-perturbative scheme to control the UV divergence.

The one-loop photon polarization tensor on a lattice of size $N_s^3\times N_t$ reads
\begin{align}
\nonumber \Pi_{\mu\nu}(p)=&e^2\frac{1}{N_s^3N_ta^4}\sum_l{\rm Tr}\left[V_\mu\left(\frac{l}{a},\frac{l}{a}+p\right)S\left(\frac{l}{a}+p\right)  \right.\\
&\times\left. V_\nu\left(\frac{l}{a}+p,\frac{l}{a}\right)S\left(\frac{l}{a}\right) \right]
\label{polarization}
\end{align}
where the vertex functions and propagators are given in (\ref{vertex}) and (\ref{axialpropagator}). The summation over the dimensionless discrete loop momentum $l$ extends to all
roots of $z_j^{N_s}-1=0$ for spatial component and all roots of $z_4^{N_t}+1=0$ for temporal component with $z_\mu=e^{il_\mu}$.
The spatial volume of the system to be simulated is ${N_sa}^3$ and the temperature is  $T=\frac{1}{N_ta}$. In the infinite volume limit, $N_s\rightarrow\infty$, the summation over
the spatial components of $l$ becomes an integral, i. e.
\begin{equation}
\frac{1}{N_s^3}\sum_{\bf l}(...)\longrightarrow\int\frac{d^3\bf l}{(2\pi)^3}(...)
\end{equation}
with each component run from $-\pi$ to $\pi$.

Substituting (\ref{vertex}) and (\ref{axialpropagator}) into (\ref{polarization}), we obtain the explicit form of (\ref{polarization})
\begin{widetext}
\begin{align}
\nonumber \Pi_{\mu\nu}(p)=&e^2\frac{a^{-2}}{N_s^3N_t}\sum_l{\rm Tr}\left\{v_\mu\left(l+\frac{a}{2}p\right)\left[\sum_{\rho}\gamma_\rho\sin \left(l_\rho+ap_\rho+a\gamma_5\mu_5\delta_{\rho 4} \right)+{\cal M}(l+ap) \right]^{-1}\right.\\
&\times\left. v_\nu\left(l+\frac{a}{2}p\right)\left[\sum_{\sigma}\gamma_\sigma\sin \left(l_\sigma+a\gamma_5\mu_5\delta_{\sigma 4} \right)+{\cal M}(l) \right]^{-1} \right\}\label{photonselfenergy}
\end{align}
\end{widetext}
with
\begin{equation}
v_\mu(k)=\gamma_\mu\cos k_\mu+r\sin k_\mu
\end{equation}
for a general $k$.

Expanding $\Pi_{\mu\nu}$ in terms of $\mu_5$, one has
\begin{equation}
\Pi_{\mu\nu}(p)=\mu_5{\Pi}^{(1)}_{\mu\nu}(p)+{\cal O}(\mu_5^3)
\end{equation}
with
\begin{widetext}
\begin{align}
\nonumber &{\Pi}^{(1)}_{\mu\nu}(p)=\left.\frac{\partial}{\partial \mu_5}\Pi_{\mu\nu}(p)\right|_{\mu_5=0}=-e^2\frac{a^{-1}}{N_s^3N_t}\sum_l\frac{1}{\sin^2(l+ap)+{\cal M}^2(l+ap)}\frac{1}{\sin^2 l+{\cal M}^2 (l)}\\
\nonumber&\times{\rm Tr}\left\{\frac{v_\mu\left(l+\frac{a}{2}p\right)\left[{\cal M}(l+ap)- \gamma\cdot \sin(l+ap) \right]\gamma_4\gamma_5\cos(l_4+ap_4)
\left[{\cal M}(l+ap)- \gamma\cdot \sin(l+ap) \right]v_\nu\left(l+\frac{a}{2}p\right)\left[{\cal M}(l)-\gamma\cdot\sin l \right]}
{\sin^2(l+ap)+{\cal M}^2(l+ap)}\right.\\  &+\left.\frac{v_\mu\left(l+\frac{a}{2}p\right)\left[{\cal M}(l+ap)- \gamma\cdot \sin(l+ap) \right]v_\nu\left(l+\frac{a}{2}p\right)\left[{\cal M}(l)- \gamma\cdot \sin(l) \right]\gamma_4\gamma_5\cos l_4\left[{\cal M}(l)- \gamma\cdot \sin(l) \right]}
{\sin^2(l)+{\cal M}^2(l)}\right\}
\label{triangle}
\end{align}
\end{widetext}

An expansion of the integrand into powers of the lattice spacing $a$ is straightforward and does not encounter an infrared divergence. Technically, all terms odd in $\sin l_\mu$ do
not contribute and the summation over the loop momentum $l$ should be symmetric under permutation of its spatial components. The final expression reads
\begin{equation}
\Pi_{ij}^{(1)}(p)={\cal I}e^2\sum_k\epsilon_{ikj}p_k+{\cal O}(a)\label{selfenergy}
\end{equation}
with
\begin{equation}
{\cal I}= 12\frac{1}{N_s^3N_t}\sum_l\frac{{\cal N}(l)}{\left[ \sin ^2l+{\cal M}^2(l)\right]^3}
\end{equation}
where
\begin{align}
\nonumber {\cal N}(l)=&\left[\sin^2 l_3-\sin^2 l_4 -{\cal M}^2(l)\right]\cos l_1\cos l_2\cos l_3\cos l_4\\
&+ 4r {\cal M}(l)\cos l_1\cos l_2\sin^2 l_3\cos l_4
\end{align}
For different lattice size, the value of $\cal I$ can be calculated numerically and the results had been listed in the Table I.
In the numerical calculation, the Wilson parameter had been fixed as $r=1$ and the mass $aM=am+4r$ had been fixed via the hopping parameter $\kappa\equiv 1/2aM=0.1665$.

In the limit $N_s\to\infty$, we have
\begin{equation}
{\cal I}= 12 \frac{1}{N_t}\sum_{l_4}\int\frac{d^3 { \textit{\bf l}}}{(2\pi)^3}\frac{{\cal N}(l)}{\left[ \sin ^2l+{\cal M}^2(l)\right]^3}
\end{equation}
Using the following identity\cite{KarstenSmit},
\begin{align}
\nonumber &{\cal M}^2(l)\cos l_\beta-4r{\cal M}(l)\sin^2 l_\beta=\cos l_\beta\left({\cal M}^2(l)+4\sin^2 l_\beta \right)\\
&+\left({\cal M}^2(l)+\sin^2 l\right)^3\sin l_\beta\frac{\partial}{\partial l_\beta}\left({\cal M}^2(l)+\sin^2 l\right)^{-2}\label{KarstenSmitidentity}
\end{align}
and integrating the spatial loop momentum ${ \textit{\bf l}}$ by part, one ends up with
\begin{align}
\nonumber {\cal I}= & 12 e^2\frac{1}{N_t}\sum_{l_4}\int\frac{d^3 { \textit{\bf l}}}{(2\pi)^3}\frac{\cos l_1\cos l_2\cos l_3\cos l_4 }{\left[ \sin ^2l+{\cal M}^2(l)\right]^3}\\
&\times\left(2\sin^2l_3-\sin^2 l_1-\sin^2l_2 \right)\label{coefficientintegral}
\end{align}
Obviously, this result is zero as long as there is no infrared singularity in the integrand, which is the case when the mass is nonvanishing.
On the other hand, at finite temperature the zeroth Mastubara fermion frequency plays a natural role as an infrared cutoff even for the massless fermions and thus this result is zero again.

\begin{table}
\caption{\label{tab:table}Numerical values of $\cal I$ in equation (\ref{selfenergy}) for several different lattice size. }
\begin{ruledtabular}
\begin{tabular}{cc}
Lattice size & $\cal I$ \\
\hline
$N_s=6,N_t=4$& $1.347\times 10^{-2}$ \\
$N_s=12,N_t=4$&  $2.439\times 10^{-4}$  \\
$N_s=20,N_t=4$&  $8.886\times 10^{-7}$\\
$N_s=50,N_t=8$&  $4.512\times 10^{-9}$\\
\end{tabular}
\end{ruledtabular}
\end{table}

The way of introducing the axial chemical potential on a lattice is not unique. An alternative approach is to insert the $U_A(1)$ link into the Wislon term as well, i.e.,
to replace the $r$-term of (\ref{yamamotoaction}) by
\begin{align}
\nonumber \frac{r}{2a}\sum_\mu\left[{\bar{\psi}}(x)\exp\{ia[ev_\mu(x)+\gamma_5 a_\mu(x)]\}\psi(x+a_\mu)\right.\\
+\left.{\bar{\psi}}(x+a_\mu)\exp\{-ia[ev_\mu(x)+\gamma_5 a_\mu(x)]\}\psi(x)\right]
\label{alternative}
\end{align}
Consequently, the mass term in the propagator (\ref {axialpropagator}) becomes
\begin{equation}
{\cal M}_c(ap)=r\sum_{\mu}\left[1-\cos a(p_\mu+\gamma_5\mu_5\delta_{\mu4})\right]
\end{equation}
and the factor $\gamma_4\gamma_5\cos(...)$ in (\ref{triangle}) is replaced by
\begin{equation}
\gamma_4\gamma_5\cos(...)+\gamma_5\sin(...)
\end{equation}
with ... equal to $l_4+ap_4$ or $l_4$. Consider the contribution of the sine term above to one of the terms of (\ref{triangle}), say the second term where the
axial vector vertex $\gamma_4\gamma_5\cos l$ is replaced by $\gamma_4\gamma_5\cos l+\gamma_5\sin l$. For a nonvanishing trace, we need a product of four gamma matrices
from rest part of the term. This can be obtained by pulling out a sine term from one of the vector vertices or a mass term from one of the propagators together
with the gamma matrices from other vector vertex and propagators. Because of the duplicated propagators on both side of the axial vector vertex, the trace
vanishes unless we pulling out the mass term ${\cal M}(l)$ from either on of the duplicated propagators. But the trace of the two terms thus obtained cancel
each other and the contribution of the sine term pertaining to the axial vector vertex is zero. The same is true for the other term in (\ref{triangle}) and our result
after (\ref{triangle}) are intact with this alternative formulation of the axial chemical potential.

\section{The case of zero temperature}

In the previous section, we investigated CME by calculating the one-loop photon polarization tensor, the derivative of which with respect to the axial
chemical potential is equivalent to the usual triangle diagrams with zero energy-momentum flowing in their axial vector vertex,
which corresponds to CME with a constant axial chemical potential. To be specific, the limit of zero energy-momentum $\lim_{p\rightarrow 0}$ of the CME
current in the previous calculations had been taken prior to $\lim_{T\to 0}$. The nonzero fermionic Matsubara energy prevents the denominator of the integrand of
(\ref{coefficientintegral}) from vanishing even in the chiral limit and zero chiral magnetic current emerges. In this section, we shall reverse the order of limit
by retaining the $p$-dependence in the denominator while taking the limit $N_t\to\infty$ and $N_s\to\infty$. All components of the loop momentum becomes
continuous and the summation becomes integral over the Brillouin zone.

Notice that, the photon polarization tensor can be related to the axial anomaly by taking the following limit
\begin{align}
\nonumber \Pi_{ij}(q)&\equiv \Lambda_{ij4}(q)\\
&=-\lim_{q_4\rightarrow 0}\frac{1}{q_4}\sum_{\rho}\frac{2}{a}\sin\frac{1}{2}a(Q_1+Q_2)_\rho\Lambda_{ij\rho}(Q_1,Q_2)
\label{anomaly}
\end{align}
where we denote the general amplitude of triangle diagrams by $\Lambda_{ij\rho}(Q_1,Q_2)$ with $Q_1\equiv ({\bf q}, q_4/2)$ and $Q_2\equiv(-{\bf q}, q_4/2)$.
Naively, this relation renders a nonvanishing CME current because of the chiral anomaly, which holds to all orders of interaction at arbitrary temperature
and chemical potential, and thus appears contradicting to the finite temperature conclusion that we obtained in previous section. But the limiting procedure
in (\ref{anomaly}) is beyond the Matsubara formulation because of the discreteness of $q_4$.

To show (\ref{anomaly}) explicitly, we notice that the amplitude of one-loop triangle diagrams reads
\begin{widetext}
\begin{align}
\nonumber \Lambda_{ij\rho}(Q_1,Q_2)=-& e^2\int\frac{d^4l}{(2\pi)^4}{\rm Tr}[\gamma_5 V_\rho\left(l+Q_1,l-Q_2\right)S(l+Q_1)V_i(l,l+Q_1)S(l)V_j(l,l-Q_2)S(l-Q_2) \\
+& \gamma_5 V_\rho\left(l+Q_2,l-Q_1\right)S(l+Q_2)V_j(l,l+Q_2)S(l)V_i(l,l-Q_1)S(l-Q_1)]
\end{align}
\end{widetext}
Its divergence with respect to the axial vertex reads
\begin{widetext}
\begin{align}
\nonumber &\sum_{\rho}\frac{2}{a}\sin\frac{1}{2}a(Q_1+Q_2)_\rho\Lambda_{ij\rho}(Q_1,Q_2)\\
\nonumber=&- e^2\int\frac{d^4l}{(2\pi)^4}{\rm Tr}\left\{\gamma_5\sum_{\rho}\frac{2}{a}\sin\frac{1}{2}a(Q_1+Q_2)_\rho\Gamma_\rho\left[al+\frac{1}{2}(aQ_1-aQ_2)\right]\right.\\
&\times \left. \left[S(l+Q_1)v_i(al+\frac{1}{2}aQ_1)S(l)v_j(al-\frac{1}{2}aQ_2)S(l-Q_2)+S(l+Q_2)v_j(al+\frac{1}{2}aQ_2)S(l)v_i(al-\frac{1}{2}aQ_1)S(l-Q_1) \right]\right\}
\end{align}
with $\Gamma_\rho(p)=\gamma_\rho\cos p_\rho$. Using the identity,
\begin{align}
 \sum_{\rho}\frac{2}{a}\sin\frac{1}{2}a(Q_1+Q_2)_\rho\Gamma_\rho\left[al+\frac{1}{2}(aQ_1-aQ_2)\right]
= S^{-1}(l+Q_1a)-{\cal M}(l+Q_1a)-S^{-1}(l-Q_2a)+{\cal M}(l-Q_2a)
\end{align}
one has
\begin{equation}
\sum_{\rho}\frac{2}{a}\sin\frac{1}{2}a(Q_1+Q_2)_\rho\Lambda_{ij\rho}(Q_1,Q_2)={\cal G}^{(1)}_{ij}(Q_1,Q_2)+{\cal G}^{(2)}_{ij}(Q_1,Q_2)
\end{equation}
with
\begin{equation}
{\cal G}^{(1)}_{ij}(Q_1,Q_2)=(Q_1+Q_2)e^2\int\frac{d^4l}{(2\pi)^4}\frac{\partial}{\partial l_4}{\rm Tr}\left[\gamma_5v_i(al+\frac{1}{2}aQ_1)S(l)v_j(al+\frac{1}{2}aQ_1)S(l+Q_1) \right]\label{partialintegral}
\end{equation}
and
\begin{align}
\nonumber {\cal G}^{(2)}_{ij}(Q_1,Q_2)=&-2(Q_1+Q_2)e^2\int\frac{d^4l}{(2\pi)^4}\left[\frac{{\cal M}(al+aQ_1)\cos a(l+Q_1)_4}{\sin^2 a(l+Q_1)+{\cal M}^2(al+aQ_1)}+\frac{{\cal M}(al)\cos al_4}{\sin^2 al+{\cal M}^2(al)} \right]\\
&\times {\rm Tr}\left[\gamma_5\gamma_4v_i(al+\frac{1}{2}aQ_1)S(l)v_j(al+\frac{1}{2}aQ_1)S(l+Q_1)\right]
\end{align}

${\cal G}^{(1)}_{ij}$ vanishes because its integrand is a total derivative and we are left with
\begin{equation}
\Pi_{ij}(q)=\frac{1}{Q_1+Q_2}{\cal G}_{ij}^{(2)}=-16e^2\sum_k\epsilon_{ikj}q_k\int\frac{d^4 l}{(2\pi)^4}\cos l_1\cos l_2\cos l_4\frac{{\cal M}^2(l)\cos l_3-4r{\cal M}(l)\sin^2 l_3}{[\sin^2 l+{\cal M}^2(l)]^3}
\end{equation}
\end{widetext}
This is exactly the chiral anomaly obtained by Karsten and Smit\cite{KarstenSmit}. Using the identity (\ref{KarstenSmitidentity}), one obtains
\begin{equation}
\Pi_{ij}(q)=\frac{e^2}{2\pi^2}\sum_k\epsilon_{ijk}q_k
\end{equation}

The analysis in this section applies also to the situation at a nonzero temperature $T$ but with the external momentum $T<<p<<\frac{1}{a}$.

\section{Concluding Remarks}

To summarize, we have calculated the chiral magnetic current to one-loop order with a lattice regularization with Wilson fermions and reproduced 
the one-loop result in continuum with the Pauli-Villars regularization. 
The Wilson fermion formulation was employed because it is analytically tractable and was adopted in the full QCD simulation reported in Ref.\cite{Yamamoto}, 
which motivated this research. For a lattice of size $N_s^3\times N_t$ with an isotropic lattice spacing, we found that the chiral magnetic current 
vanishes in the limit $N_s\to\infty$ at isotropic lattice spacing $a$ but shows up when the limits $N_t$ and $N_s\to\infty$ are taken simultaneously. 
In the continuum limit $a\to 0$, both $N_t, N_s\to\infty$ and the null result corresponds to $N_t/N_s\to 0$. Consequently, there is no 
chiral magnetic effect at  nonzero temperature in thermal equilibrium to one-loop order. Our conclusion is also consistent with the one-loop numerical result of the overlapping
fermions\cite{Buividovich2}.

To assess the finite size effect, we have calculated numerically the one-loop chiral magnetic current on a lattice 
with the same parameters as that of the full QCD simulation reported in \cite{Yamamoto} and found a nonzero chiral magnetic current with a very small coefficient $C$ compared to the 
that of the classical formula, i.e $C=\frac{1}{2\pi^2}\simeq 0.05$. A sizable coefficient $C\simeq 0.013$ was extracted from the full QCD simulation reported in \cite{Yamamoto}. 
In a follow up work\cite{Yamamoto1}, the same author calculated the CME current on the same size lattice and a larger one ($N_s=18$ and $N_t=6,10,18$) but under the quenched approximation, and estimated 
that $C_{\rm latt.}\simeq0.02$ to $0.03$. In our opinion, it is still premature to attribute this value to the QCD dynamics or to the finite size effect beyond one-loop order.
As was shown nonperturbatively in\cite{HouRen}, the constant limit of a spacetime dependent axial chemical potential is rather subtle. 
If the homogeneity is switched off prior to the time dependence, the classical chiral magnetic current emerges because of the anomalous Ward idensity. 
But this order of the limit goes beyond the Mastubara formulation underlying the thermal equilibrium. 
If the homogeneity limit is taken after static limit at a nonzero temperature, the link to the anomaly is broken and the chiral magnetic current 
vanishes because of the electromagnetic $U(1)$ gauge invariance, in the \textit{absence} of infrared divergences from higher orders. On the other hand, the infrared problems of 
the thermal QCD is rather serious. A simple power counting argument in analoguous to that in 
\cite{Linde} applied to the term of a thermal diagram with zero Matsubara frequency in all gluon internal lines shows that this term is proportional to $(T^2/q^2)^n$ 
with the power $n>0$ increasing with the number of 3-gluon and 4-gluon vertices. Such an infrared catastrophe may be 
regulated by the nonperturbative magnetic mass, which was estimated to be $m_M\simeq 5.8T$ \cite{Maezawa} for two flavors. The continuum limit requires that $m_Ma<<1$ and 
the infinite size limit requires that $m_ML_s>>1$. The Tab. 2 below displays the parameters $m_Ma$ and $m_ML_s$ of the lattices examined in\cite{Yamamoto1} with quenched
simulations. While the continuum limit appears marginal, the spatial size of the lattice seems fairly large gauged on the measured magnetic mass. 
But the extracted CME coefficient $C$ does not show significant 
drop with increasing temperature if the nonzero $C$ is an artifact of finite size effect. It would be interesting to examine the tendency of $C$ with an increasing spatial size 
of the lattice at a fixed temperature and with a smaller $m_Ma$.

\begin{table}
\caption{\label{tab:table2} The lattice parameters of the quenched simulation in \cite{Yamamoto1} regarding the continuum limit and finite size effect.}
\begin{ruledtabular}
\begin{tabular}{cccccccc}
$N_s$ & $N_t$ & $a(\rm{GeV}^{-1})$ & $L_s(\rm{GeV}^{-1})$ & $T(\rm{MeV})$ & $m_M(\rm{MeV})$ & $m_Ma$ & $m_ML_s$ \\
\hline
12 & 4 & 0.53 & 6.3 & 475 & 2760 & 1.45 & 17.4 \\ 
12 & 6 & 0.53 & 6.3 & 317 & 1840 & 0.97 & 11.6 \\
12 & 12 & 0.53 & 6.3 & 158 & 918 & 0.48 & 5.8 \\
18 & 6 & 0.32 & 5.8 & 517 & 3000 & 0.97 & 17.4 \\
18 & 10 & 0.32 & 5.8 & 310 & 1800 & 0.58 & 10.44 \\
18 & 18 & 0.32 & 58 & 172 & 1000 & 0.32 & 5.8\\
\end{tabular}
\end{ruledtabular}
\end{table}


\begin{acknowledgments}
This research is partly supported by the
Ministry of Science and Technology of China (MSTC) under the "973" Project No. 2015CB856904(4).  B. F. is supported by NSFC under grant No. 11305067, No. 11535005. D-f. Hou and H-c Ren are partly supported by NSFC under Grant Nos. 11375070, 11521064.
\end{acknowledgments}



\newpage 


\begin{thebibliography}{}

\bibitem{CME1}D. E. Kharzeev, L. D. McLerran and H. J. Warringa, \textit{The effects of topological charge change in heavy ion collisions:'Event by event $\cal{P}$ and $\cal{CP}$ violation',} Nucl. Phys. \textbf{A 803}, 227 (2008).

\bibitem{CME2}K. Fukushima, D. E. Kharzeev and H. J. Warringa, \textit{The chiral magnetic effect,} Phys. Rev. D \textbf{78},  074033 (2008).

\bibitem{CME3}Y. Burnier, D. E. Kharzeev, J-f Liao and H-U Yee, \textit{Chiral magnetic wave at finite baryon density and the electric quadrupole moment of quark-gluon plasma in heavy ion collisions,} Phys. Rev. Lett., \textbf{107},  052303 (2011).

\bibitem{Weylsemi1}Q. Li et al., \textit{Chiral magnetic effect in $ZrTe_5$,} Nature Physics \textbf{12}, 550 (2016).

\bibitem{Weylsemi2}H. J. Kim et al., \textit{Dirac versus Weyl Fermions in topological insulators: Adler-Bell-Jackiw anomaly in transport phenomena,} Phys. Rev. Lett. \textbf{111}, 246603 (2013).

\bibitem{Weylsemi3}J. Xiong et al., \textit{Evidence for the chiral anomaly in the Dirac semimetal $Na_3Bi$,} Science \textbf{350}, 413 (2015).

\bibitem{Weylsemi4}X-c. Huang et al., \textit{Observation of the Chiral-anomaly-induced negative magnetoresistance in 3D Weyl semimetal $TaAs$,} Phys. Rev. X \textbf{5}, 031023 (2015).

\bibitem{NN1}H. B. Nielsen and M. Ninomiya, \textit{The Adler-Bell-Jackiw anomaly and Weyl fermions in a crystal,} Phys. Lett. \textbf{B 130}, 389 (1983).

\bibitem{tHooft1}G. 't Hooft, \textit{How instantons solve the $U(1)$ problem,} Phys. Rept. \textbf{142}, 357 (1986).

\bibitem{tHooft2} G. 't Hooft, \textit{Computation of the quantum effects due to a four-dimensional pseudoparticle,} Phys. Rev. D \textbf{14} 3432 (1976).[Erratum Phys. Rev. D \textbf{18}, 2199 (1978).]

\bibitem{JackiwRebbi}R. Jackiw and C. Rebbi, \textit{Vacuum periodicity in a Yang-Mills quantum theory,} Phys. Rev. Lett. \textbf{37}, 172 (1976).

\bibitem{CallanGross}C. G. Callan, R. F. Dashen and D. J. Gross, \textit{The structure of the gauge theory vacuum,} Phys. Lett. \textbf{B 63}, 334 (1976).

\bibitem{Kharzeev}Y. Hirono, D. E. Kharzeev and Y. Yin, \textit{Quantized chiral magnetic current from reconnections of magnetic flux,} Phys. Rev. Lett. \textbf{117}, 172301 (2016).

\bibitem{CMEinRHIC1}B. I. Abelev et al., \textit{Azimuthal charged particle correlations and possible local strong parity violation,} Phys. Rev. Lett. \textbf{103}, 251601 (2009).

\bibitem{CMEinRHIC2}L. Adamczyk et al., \textit{Observation of charge asymmetry dependence of pion ellipic flow and the possible chiral magnetic wave in heavy-ion collisions,} Phys. Rev. Lett. \textbf{114}, 252302 (2015).

\bibitem{CMEinAlice}R. Belmont, \textit{Charge-dependent anisotropic flow studies and the search for the chiral magnetic wave in ALICE,} Nucl Phys. \textbf{A 931}, 981 (2014).

\bibitem{CMEtaskforce}V. Skokov et al., \textit{Chiral magnetic effect task force report,} arXiv:1608.00982[nucl-th].

\bibitem{rubakov}V. A. Rubakov, \textit{On chiral magnetic effect and holography,} arXiv:1005.1888[hep-ph].

\bibitem{HouRen}D-f Hou, H. Lui and H-c Ren, \textit{Some field theoretic issues regarding the chiral magnetic effect,} JHEP \textbf{05}, 046 (2011).

\bibitem{Buividovich1}P. V. Buividovich, M. N. Chernodub, E. V. Luschevskaya and M. I. Polikarpov, \textit{Numerical evidence of chiral magnetic effect in lattice gauge theory,} Phys. Rev. D \textbf{80}, 054503 (2009).

\bibitem{Abramczyk}M. Abramczyk, T. Blum, G. Petropoulos and R. Zhou, \textit{Chiral magnetic effect in $2+1$ flavor QCD$+$QED,} PoS LAT \textbf{2009}, 181 (2009).

\bibitem{Yamamoto}A. Yamamoto, \textit{Chiral magnetic effect in lattice QCD with a chiral chemical potential,} Phys. Rev. Lett. \textbf{107}, 031601 (2011).

\bibitem{Yamamoto1}A. Yamamoto, \textit{Lattice study of the chiral magnetic effect in a chirally imbalanced matter} Phys. Rev. D \textbf{84}, 114504 (2011).

\bibitem{Buividovich2}P. V. Buividovich, \textit{Anomalous transport with overlap fermions,} Nucl. Phys. \textbf{A 925}, 218 (2014).

\bibitem{Schafer}G. S. Bali, F. Bruckmann, G. Endr$\ddot{o}$di, Z. Fodor, S. D. Katz and A. Sch$\ddot{a}$fer, \textit{Local CP-violation and electric charge separation by magnetic fields from lattice QCD,} JHEP \textbf{1404}, 129 (2014).

\bibitem{KarstenSmit}L. H. Karsten and J. Smit, \textit{Lattice fermions: species doubling, chiral invariance and the triangle anomaly,} Nucl. Phys. \textbf{B 183}, 103 (1981).

\bibitem{NN2}H. B. Nielsen and M. Ninomiya, \textit{A no-go theorem for regularizing chiral fermions,} Phys. Lett. \textbf{B 105}, 219 (1981).

\bibitem{KharzeevandWarringa}D. E. Kharzeev and H. J. Warringa, \textit{Chiral magnetic conductivity,} Phys. Rev. D \textbf{80}, 034028 (2009).

\bibitem{Landsteiner}I. Amado, K. Landsteiner and F. Pena-Benitez, \textit{Anomalous transport coefficients from Kubo formulas in Holography,} JHEP \textbf{1105}, 081
(2011).



\bibitem{Linde}A. Linde, \textit{Infrared problem in thermodynamics of the Yang-Mills gas}, Phys. Lett. \textbf{B 89}, 289 (1980).

\bibitem{Maezawa}Y. Maezawa, et. al., \textit{Electric and magnetic screening massesa at finite temperature from generalized Polyakov-line correlations in two-flavor
lattice QCD}, Phys. Rev. D \textbf{81}, 091501 (2010).















\end{thebibliography}
\end{document}